# HMACA: Towards Proposing a Cellular Automata Based Tool for Protein Coding, Promoter Region Identification and Protein Structure Prediction

Prof Pokkuluri Kiran Sree[1], Dr. Inampudi Ramesh Babu[2] and
Smt S.S.S.N. Usha Devi Nedunuri[3]

[1]Professor, Department of C.S.E, BVC Engg College, Odalarevu, India
[2] Professor, Department of Computer Science & Engineering, Acharya Nagarjuna University
[3] Assistant Professor, Dept of CSE, Jawaharlal Nehru Technological Universtiy, Kakinada

*ABSTRACT*

*Human body consists of lot of cells, each cell consist of DeOxaRibo Nucleic Acid (DNA). Identifying the genes from the DNA sequences is a very difficult task. But identifying the coding regions is more complex task compared to the former. Identifying the protein which occupy little place in genes is a really challenging issue. For understating the genes coding region analysis plays an important role. Proteins are molecules with macro structure that are responsible for a wide range of vital biochemical functions, which includes acting as oxygen, cell signaling, antibody production, nutrient transport and building up muscle fibers. Promoter region identification and protein structure prediction has gained a remarkable attention in recent years. Even though there are some identification techniques addressing this problem, the approximate accuracy in identifying the promoter region is closely 68% to 72%. We have developed a Cellular Automata based tool build with hybrid multiple attractor cellular automata (HMACA) classifier for protein coding region, promoter region identification and protein structure prediction which predicts the protein and promoter regions with an accuracy of 76%. This tool also predicts the structure of protein with an accuracy of 80%.*

## I. INTRODUCTION

Mathematical computing can be applied to most problems in biology. In bioinformatics the techniques of computer algorism are used to examine the information available with the bimolecules of highest order. Bioinformatics consists of how to store data, presenting the feature within the data and retrieval of the data also. Promoters are molecules with macro region that are responsible for a wide range of vital biochemical functions, which includes acting as oxygen, nutrient transport and building up muscle fibers. Specifically, the Promoters are chains of amino acids and DNA sequences, of which there are 20 different types, coupled by peptide bonds [2]. The structural hierarchy possessed by Promoters is typically referred to as primary and tertiary region. Promoter Region Predication from sequences of amino acid gives tremendous value to biological community.

## II. RELATED WORK

Reese MG al [2] has proposed a Neural Network Model for predicting the promoter region. Steen Knudsen al [3] has used statistical classifiers to identify promoter regions. Techniques for region identification include, but are not limited to, constraint programming methods, statistical approaches to predict the probability of an amino acid being in one of the structural elements, and Bayesian network models. The Objective of structure prediction is to identify whether the amino acid residue of protein is in helix, strand or any other shape. In 1960 as a initiative step of structure prediction the probability of respective structure element is calculated for each amino acid by taking single amino acid properties consideration [1],[3],[6] . The third generation technique includes machine learning, knowledge about proteins, several algorithms which gives 70% accuracy. Neural Networks[10],[11] are also useful in implementing structure prediction programs like PHD, SAM-T99.

## III. HYBRID MULTIPLE ATTRACTOR CELLULAR AUTOMATA (HMACA)

The linear/additive HMACA are amenable to detailed characterization with linear algebraic tools. Due to the absence of such a mathematical tool, there has been varied effort with different parameters to characterize non-linear HMACA .We detail the characterization of each of the categories separately. However, some very interesting works simulating non-linear CA from product of linear CA are recently reported in [3]. These works are aimed at taking the advantage of linear algebraic tools to characterize the wide variety of non-linear CA state transition. One of the major thrust has been to study the non-linear CA dynamics as it evolves in successive time steps. The emergent patterns in the decentralized systems give rise to some form of globally coordinated behavior. A detailed study of CA dynamics helps us to understand the emergent behavior and analyze its computational power [1, 10].

CA classification based on the study of its dynamics was a major interest for the researchers. Borrowing the concept from the field of continuous dynamical systems, Wolfram [9] first classified CA into four broad categories:

- Class 1: CA which evolve to a homogeneous state;
- Class 2: those which evolve to simple separated periodic structures;
- Class 3: which exhibit chaotic or pseudo-random behavior; and
- Class 4: the class of CA displaying complex patterns of localized structures and are capable of universal computation [9].

### 3.1 Population Generation

**Algorithm**

Input: Pattern set P to be memorized, Maximum Generation ($G_{max}$).
Output: Dependency String (DES) and associated information.
    begin

Step 1: Generate 500 new chromosomes for initial population (IP1).
Step 2: Initialize generation counter GAC=zero; PP1← IP1.
Step 3: Compute fitness value F for each chromosome of PP1.
Step 4: Store DES, and corresponding information for which the fitness value F = 100%.
Step 5: If F = 100% for at least one chromosome of PP1, then go to Step 12.
Step 6: Rank chromosomes in order of fitness.
Step 7: Increment generation counter (GAC)
Step 8: If GAC > $G_{max}$ then go to Step 11.
Step 9: Form NP by selection, crossover and mutation.
Step 10: PP1← NP; Go to Step 3.
Step 11: Store DS, and corresponding information for which fitness value is maximum.
Step 12: Stop.

### 3.2 HMACA Tree Building

Input : Training set S = {S1, S2, ··, SK}
Output : HMACA Tree.
Partition(S, K)
Step 1 : Generate a HMACA with k number of attractor basins.
Step 2 : Distribute S into k attractor basins (nodes).
Step 3 : Evaluate the distribution of examples in each attractor basin
Step 4 : If all the examples (S') of an attractor basin (node) belong to only one class, then label the attractor basin.
Step 5 : If examples (S') of an attractor basin belong to K' number of classes, then, Partition (S', K').
Step 6 : Stop.

## IV. EXPERIMENTAL RESULTS

We have conducted experiments on ENCODE datasets and FICKETT &TOUNG data sets. The proposed interface is shown in Figure 1. The sample outputs and accuracies are also reported in Table 1.

**Figure 1. Proposed Interface**

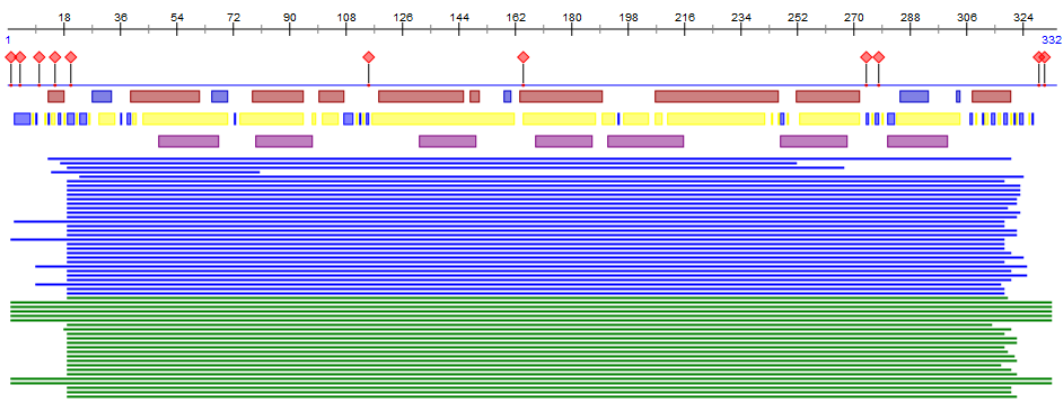

**Sample Output**

```
Seq-Pos-Residue            ANN     HMM     NES     Predicted
#-------------------------------------------------------
Sequence-1-A               0.098   0.000   0.000   -
Sequence-2-S               0.089   0.000   0.000   -
Sequence-3-Q               0.099   0.000   0.000   -
Sequence-4-K               0.095   0.000   0.000   -
Sequence-5-R               0.092   0.000   0.000   -
Sequence-6-P               0.080   0.000   0.000   -
Sequence-7-S               0.082   0.000   0.000   -
Sequence-8-Q               0.080   0.000   0.000   -
Sequence-9-R               0.077   0.000   0.000   -
Sequence-10-H              0.084   0.000   0.000   -
Sequence-11-G              0.094   0.000   0.000   -
Sequence-12-S              0.082   0.000   0.000   -
Sequence-13-K              0.070   0.000   0.000   -
Sequence-14-Y              0.077   0.000   0.000   -
Sequence-15-L              0.129   0.000   0.000   -
Sequence-16-A              0.073   0.000   0.000   -
Sequence-17-T              0.095   0.000   0.000   -
Sequence-18-A              0.084   0.000   0.000   -
Sequence-19-S              0.070   0.000   0.000   -
Sequence-20-T              0.091   0.000   0.000   -
Sequence-21-M              0.093   0.000   0.000   -
Sequence-22-D              0.071   0.000   0.000   -
Sequence-23-H              0.094   0.000   0.000   -
Sequence-24-A              0.074   0.000   0.000   -
Sequence-25-R              0.087   0.000   0.000   -
```

Table 1. Accuracies Reported

| Prediction Method | Prediction Accuracy for Protien | Prediction Accuracy for Promoter | Prediction Accuracy Protein Structure Prediction |
|---|---|---|---|
| DSP | 62% | 70% | 66% |
| PHD | 70% | 68% | 74% |
| SAM-T99 | 68% | 77% | 77% |
| SS Pro | 70% | 73% | 81% |
| HMACA | 75% | 85% | 97% |

## V. CONCLUSION

HMACA predicts the protein coding regions from DNA sequence and provides the best overall accuracy that ranges between 77% and 88.7%. To provide a more thorough analysis of the viability of our proposed technique many experiments were conducted. Our extensive results indicate that such a level of accuracy is attainable, and can be potentially surpassed with our method. HMACA predicts the structure of protein with an accuracy of 84% and promoter identification with an accuracy of 76%.